\documentclass[journal]{IEEEtran}

\usepackage{cite}      

\usepackage{graphicx}  

\usepackage{psfrag}    

\usepackage{subfigure} 
\begin{document}
%
\title{Gas Purity effect on GEM Performance in He and Ne at Low Temperatures}
%
%
\author{ R.~Galea$^*$\thanks{$^*$Corresponding author
  galea@nevis.columbia.edu.}, J.~Dodd, Y.~Ju, M.~Leltchouk,
  W.~Willis: Nevis 
  Laboratories, Columbia University, Irvington, NY 10533, USA\\
 P.~Rehak, V.~Tcherniatine: Brookhaven National Laboratory, Upton, NY
  11973, USA \\
A.~Buzulutskov, D.~Pavlyuchenko: Budker Institute for Nuclear Physics,
  Novosibirsk 630090, Russia}

\maketitle

\begin{abstract}
The performance of Gas Electron Multipliers (GEMs) in gaseous He, Ne,
He+H$_2$ and Ne+H$_2$ was studied at temperatures in the range of 3-293~K. 
This paper reports on previously published measurements and additional
studies on the  
effects of the purity of the gases in which the GEM performance is evaluated.
In He, at temperatures between 77 and 293~K, triple-GEM structures
operate at rather high gains, exceeding 1000. There is an
indication that this high gain is achieved through the Penning effect
as a result of impurities in the gas. At lower temperatures the
gain-voltage characteristics are significantly modified probably due
to the freeze-out of these impurities. Double-GEM and single-GEM
structures can operate down to 3~K at gains reaching only several tens
at a gas density of about 0.5~g/l; at higher densities the maximum
gain drops further. In Ne, the maximum gain also drops at cryogenic
temperatures. The gain drop in Ne at low temperatures can be
re-established in Penning mixtures of Ne+H$_2$: very high gains, exceeding
10$^4$, have been obtained in these mixtures at 30-77~K, at a density of
9.2~g/l which corresponds to saturated Ne vapor density at 27~K. The
addition of small amounts of H$_2$ in He also re-establishes large GEM
gains   
above 30~K but no gain was observed in He+H$_2$ at 4~K and a density of
1.7~g/l (corresponding to roughly one-tenth of the saturated vapor
density). These studies are, in part, being pursued in the development 
of two-phase He and Ne detectors for solar neutrino detection.
\end{abstract}


\section{Introduction}
%
%
%
%
\PARstart{T}{he} motivation for studying the performance of Gas 
Electron Multipliers (GEMs) at cryogenic temperatures stems from 
the eBubble project~\cite{nevisebubble}, which aims 
to develop a low energy solar neutrino detector. 
The eBubble detector is currently envisioned as a two-phase detector which
 should 
operate in those liquids which allow the formation of electron 
bubbles~\cite{ebubbles}. 

The eBubble detector is a Time Projection Chamber-like (TPC) 
tracking detector. The expected signal from low energy solar
neutrino interactions comes the measurement of ionization of elastically 
scattered target electrons.
The ionization signals are expected to be small and hence the signal
 needs to be 
amplified in the saturated vapor above the liquid phase. GEMs can 
operate in noble gases at high gain~\cite{pre_gain1} and cryogenic
temperature~\cite{pre_gain2} and 
 therefore are naturally of interest. Other applications for
 this detector include but are not limited to WIMP
 detection~\cite{wimp} 
and Positron Emission Tomography~\cite{PET}.

\section{Experimental Setup}
\label{sec:setup}

\begin{figure}
\centering
\includegraphics[width=2.5in]{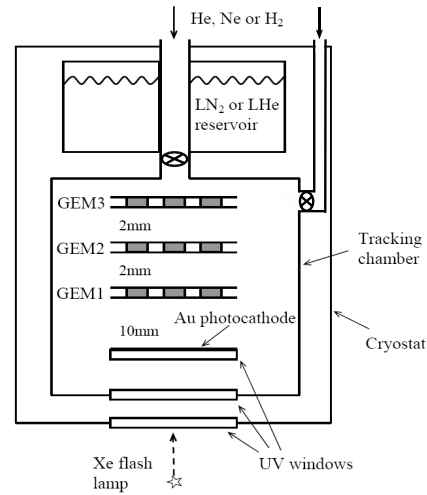}
\caption{Schematic view of the experimental setup used to study GEM
  performance at low temperatures (not to scale).}
\label{fig:setup}
\end{figure}

The experimental setup used in these studies was developed at Columbia
University Nevis
Laboratories
and at Brookhaven National Laboratory.
Figure~\ref{fig:setup} illustrates the basic components of the setup.
Three GEM foils and a photocathode were 
mounted in the tracking chamber. The chamber
was connected to a support tube which was in direct contact with  a
cryogenic liquid reservoir. This tube also served as one port
through which the gas could be filled into the chamber. A second
method of filling the chamber was 
used which sealed the central line and rather filled through a tube
which passed through the vacuum region isolated from the reservoir.

The eBubble chamber is a double walled vessel. Two copper cooling loops 
for the cryogen vapor-cooling circuits are soldered on the flanges of the chamber. 
A needle valve in the cryostat is used to supply and control the flow rate from 
the reservoir to  the
cooling circuits. 
Measurements at temperatures
between 77-293~K were performed by filling the reservoir with liquid
nitrogen and measurements at temperatures down to 2~K were performed
with liquid 
helium in the reservoir. 

The chamber was initially filled with  He, Ne and/or H$_2$ taken
directly from gas bottles with 99.999\% quoted purities. The data from
this gas supply will be henceforth  referred to as ``pure'' gas. The gas could
also be purified further by passing  the gas through
an Oxisorb\footnote{Oxisorb is a trademark of Messer Greisheim Gmbh.}
filter followed by a getter. The additional 
purification system\footnote{The additional purification system
  consisted of a SAES MonoTorr heated rare
  gas purifier. An input flow rate of 5~slpm was used which reduced
  the inlet N$_2$ impurity level to less than 1~ppb.
 For further purity performance specifications see
  {\tt http://www.puregastechnologies.com/ps4.htm}.
} is presumed to reduce the level of impurities to
1~ppb. Data from this further purified gas will be henceforth referred
to as ``ultrapure'' for clarity.

The photocathode was composed of a semitransparent gold film deposited
on a quartz substrate. A UV window at the bottom allowed the
transmission of light from a  
high-powered 
Xe flash lamp to the the photocathode. 

The GEMs were produced at CERN and had the following parameters:
50~$\mu$m thick kapton, 70 and 55~$\mu$m hole diameter on the copper
and kapton respectively, and a 140~$\mu$m hole pitch. The active area
of the GEMs was 28$\times$28~mm$^2$. The GEM foils were mounted on G10
frames with an inter-GEM separation of 2~mm. The distance between the
photocathode and the first GEM was 10~mm.

The GEM electrodes were biased through a resistive high voltage
divider placed outside the cryostat. It was comprised of three
identical circuits connected in parallel and ensured an equal voltage
drop across the GEMs and the gaps between them.

The ``calibration'' signal was induced by electrons arriving at the first
electrode of the
first GEM from the photocathode, i.e. before avalanche amplification.
Both avalanche and calibration signals were read out using a
charge-sensitive preamplifier followed by a linear amplifier.
The gain is defined as the pulse-height
of the avalanche signal at the last GEM divided by that of the calibration
signal, at a shaping time of 10~$\mu$s. The drift field was $\sim 1~$kV/cm.

\section{First measurements of GEM gain at low temperatures}

Initial measurements were performed and published~\cite{first}
in the setup described in section~\ref{sec:setup}. 
The measurements in~\cite{first} were made with gas directly taken
from gas bottles with 99.999\% quoted purity and filled through the
central tube. 

The performance of GEMs in gaseous He, Ne and mixtures of Ne and H$_2$
was studied in the temperature range of 2.6-293~K. Above 77~K in
He, the triple-GEM structures can operate at high gains exceeding
1000. The transition from LN$_2$ to LHe temperatures is characterized
by an increase in the operating voltage and a dramatic decrease of the
maximum gain. Double and single-GEM structures functioned at 2.6~K and
a density of 0.5~g/l at
a gain of order 10. At higher densities the gain drops further. At a
density of 1.6 g/l at 2.6K, the maximum gain observed in the
double-GEM is about 5. This corresponds to a gas density about an
order of magnitude less than saturated vapor density at 4.2K. At lower
temperatures 
the slope of the gain-voltage curve also decreased which indicated a
change in the avalanche mechanism.

In Ne a similar effect was observed. However, in Ne the drop in
gain could be compensated for and even enhanced by the addition of a
small fraction, $\sim 0.1\%$ of H$_2$. Very high gains, 10$^4$, were
obtained at 55~K and a 
density of 9.2~g/l which corresponds to the saturated
vapor density of Ne at 27~K. 

One possible explanation for the observed suppression of the GEM gain
at  low temperatures was the likelihood that the gain above 77~K was
a result of Penning ionization of impurities which freeze out as the
temperature is lowered to order of 4~K. 


\section{GEM Performance in Helium}

\begin{figure}
\centering
\includegraphics[width=2.5in]{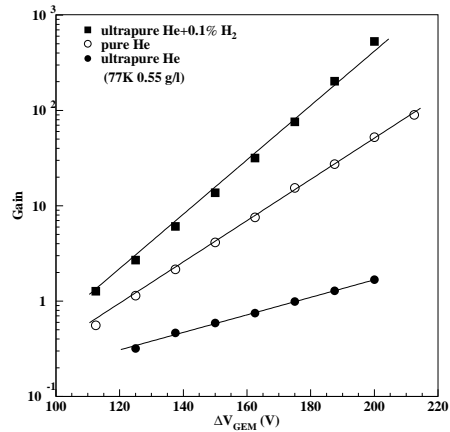}
\caption{Gain-voltage characteristics of the double-GEM in pure and
  ultrapure  He and
  He+0.1\%H$_2$ at 77~K and a density of 0.55~g/l. }
\label{helium}
\end{figure}

In order to test the impurity hypothesis, subsequent runs were
performed with additional purification of the gas source as described
in section~\ref{sec:setup}. Figure~\ref{helium} shows the gain-voltage
characteristics for a double-GEM structure at 77~K and a density of
0.55~g/l. 
There is a clear suppression of the gain in the ultrapure He as
compared to the previous data taken with pure gas directly from the gas
bottle. Also shown is the effect of the addition of 0.1\% of H$_2$
which not only restores the large gain but enhances it further. 

\begin{figure}
\centering
\includegraphics[width=2.5in]{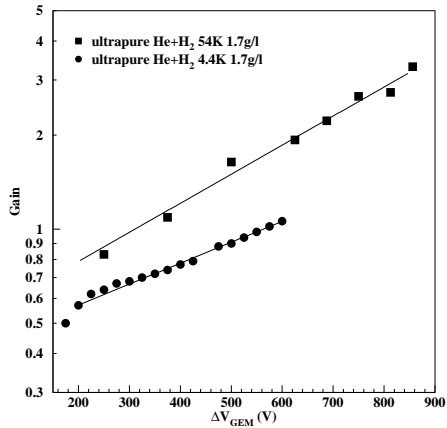}
\caption{Gain-voltage characteristics of the single-GEM in
  He+H$_2$ at a density of 1.7~g/l at 54~K and 4.4~K. }
\label{lowt_he}
\end{figure}

As the temperature is decreased to liquid helium temperatures the high
gain attained at 77~K in the He+H$_2$ mixture is lost. This is
illustrated in Fig.~\ref{lowt_he}, which shows the single-GEM
performance at a density of 1.7 g/l. 
The
increased density and the decreased effect of other mechanisms, such as
associative ionization which is presumed to decrease with
temperature~\cite{first}, could both contribute to the loss in
gain. However, at 4.4~K the H$_2$ vapor pressure is negligible and
hence any enhanced gain from Penning ionization is lost at low
temperatures. 

\section{GEM Performance in Neon}

\begin{figure}
\centering
\includegraphics[width=2.5in]{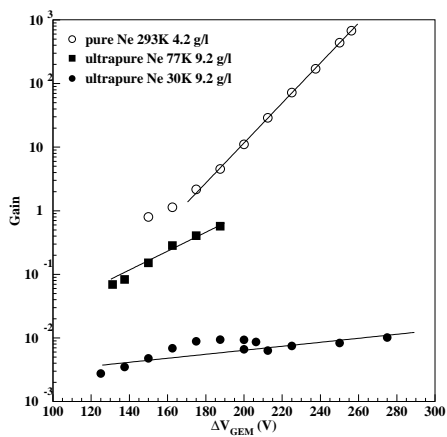}
\caption{Gain-voltage characteristics of the triple-GEM in Ne at 77~K
  and 30~K and a density of 9.2~g/l. }
\label{pureneon}
\end{figure}

The effect of purifying the source Ne gas on the GEM gain is the same
as in the case with He. Figure~\ref{pureneon} shows a comparison
between pure and ultrapure Ne gas data. As in the case between pure and
ultrapure He, the ultrapure Ne gas shows much lower gain, in a
triple-GEM, than the pure 
Ne gas above 77~K. In addition, there is a further reduction in
the signal between the 
77~K and 30~K data which is not
understood. 

\begin{figure}
\centering
\includegraphics[width=2.5in]{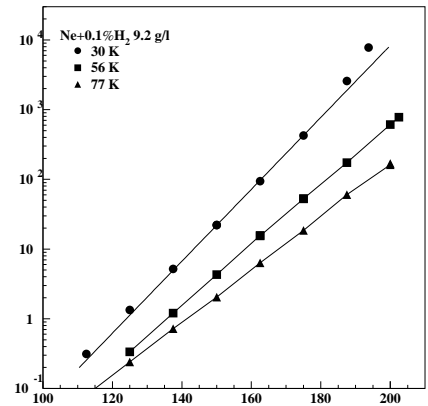}
\caption{Gain-voltage characteristics of the triple-GEM in
  ultrapure Ne+0.1\%H$_2$ at 30, 56 and 77~K at
  a density of 9.2~g/l. }
\label{negain}
\end{figure}

At low temperatures the problem of the gain drop is solved by the
addition of a small amount of H$_2$. Very high gains, 10$^4$,
are observed (see Fig.~\ref{negain}) at 30~K and a density of
9.2~g/l, which corresponds to the saturated Ne vapor density at 27~K. 
The gas mixture for all the data shown in Fig.~\ref{negain} was filled
into the chamber at 77~K and kept for all  the measurements. 
Gain curve increases with decreasing
temperature. One possible explanation for this is that the Ne vapor
pressure is falling faster than that of the H$_2$ resulting in a
different concentration of H$_2$ at the various temperature points.
Another and more likely explanation is that there is a change in the
concentration of H$_2$ as a result of stratification that can occur
due to the temperature gradient in the filling tube. 

Nonetheless, the observation of high gain in a triple-GEM  at
saturated density makes 
the Ne+H$_2$ mixture a viable option for the eBubble two-phase
detector. 

\section{Discussion}

The aim of these studies was to investigate the performance of GEMs in
ultrapure He and Ne gases at low temperatures and high densities. These are the
conditions required by an application to detect low energy solar
neutrinos~\cite{nevisebubble}. 

Previously published measurements had shown high gains in GEMs operated in He
above 62~K~\cite{first}, which decreased as the temperature was
lowered. Further measurements, presented in Fig.~\ref{helium}, show that
with 
further purification of the He a significant reduction
in the gain at 77~K occurs. This confirms the conclusion
in~\cite{first} that the gain observed above 62~K is due to Penning
ionization of impurities in the He. The addition of a controlled
impurity, namely 
H$_2$,  restored and further enhanced the gain at 77~K and
$\rho=0.55~$g/l above the aforementioned measurements. 

Figure~\ref{lowt_he} further supported the supposition that the gain
at high temperatures was a result of Penning ionization on impurities.
At 4~K in a He+H$_2$ mixture, there was no gain since the vapor
pressure of the H$_2$ is neglibile at this temperature. Some is
observed at 54~K, where the 
H$_2$ impurity affects the avalanche gain.

At 30~K and a density $\rho=9.2~$g/l, gains of order $10^4$ were
achieved in an ultrapure Ne+H$_2$ mixture. Higher gains were observed
(see Fig.~\ref{negain}) in the 
same gas mixture with decreasing temperature. This was most likely due
to a change in the H$_2$ fraction in the mixture. The H$_2$ concentration
can change as a result of
stratification that can occur due to the temperature gradient in the
filling tube, or because the relative vapor pressure of Ne was falling
faster than that of the H$_2$ at lower temperatures. 

\begin{figure}
\centering
\includegraphics[width=2.5in]{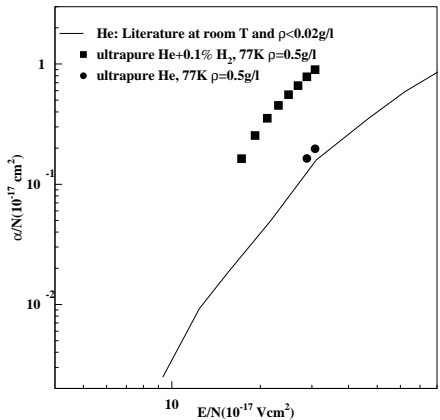}
\caption{Reduced ionization coefficients as a function of the reduced
  electric field in He and He+H$_2$. The data are
  compared to those taken from~\cite{lit1}. The
   values are calculated  from  double-GEM data.}
\label{he_alpha}
\end{figure}

\begin{figure}
\centering
\includegraphics[width=2.5in]{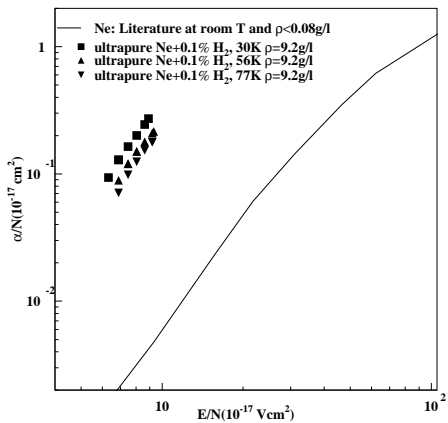}
\caption{Reduced ionization coefficients as a function of the reduced
  electric field in Ne+H$_2$. The data are
  compared to those taken from the~\cite{lit2}. The
   values are calculated  from the triple-GEM data.}
\label{ne_alpha}
\end{figure}

The measurements presented in this paper
indicate that impurities play a significant role in GEM gains in
He and Ne at temperatures above 30~K. This
conclusion is supported further by Fig.~\ref{he_alpha} and
Fig.~\ref{ne_alpha} which  show
the ionization coefficients estimated from these measurements  
according to the prescription in~\cite{associative}. Figure~\ref{he_alpha}
shows that the ionization coefficients derived from the ultrapure
He data lie close to the measurements taken from literature, where
the primary mechanism for avalanche gain is attributed to electron impact
ionization. All other data points,  derived from He+H$_2$ and
Ne+H$_2$ mixtures, show 
enhanced ionization coefficients indicating that the additional gain is due to
Penning ionization on the H$_2$ impurities.  


\section{Conclusion}

The effect of gas purity on the performance of GEMs in He and  Ne
was 
studied at low temperatures.  When the gas supply was taken
directly from gas bottles with 99.999\% quoted purities, large gains
were observed above 77~K. The gain drops significantly at lower
temperatures which is predominantly due to the freezing out of
impurities and hence the suppression of Penning ionization of
impurities. The gain was observed to decrease dramatically even
at high temperatures after
further purification of the gas supply, by additional filtering
through Oxisorb and a getter.

High gains were restored for Ne through the addition of a small amount
of H$_2$. This gain persisted at levels on the order of 10$^4$ at 30~K
and a density which corresponds to saturated vapor
density at 27~K. This makes the use of GEMs as a method of
amplification of low-energy solar neutrino signals in liquid neon an
attractive and viable option for the eBubble detector. 

\section*{Acknowledgment}
This work was supported by NSF grants PHY00-98826 and PHY05-00492.
Additional support for 
A.Buzulutskov was provided by a CRDF grant RP1-2550.



%




\end{document}